\documentclass[12pt]{iopart}

\usepackage{graphics}
\begin{document}

\title[The dispersion relation for ionacoustic instabilities in
the ionosphere at 80-200km]{
Towards the dispersion relation for ionacoustic instabilities in
weakly inhomogeneous ionospheric plasma at altitudes 80-200km and
its low-frequency solution.}

\author{O I Berngardt, A P Potekhin}

\address{
Institute of Solar-Terrestrial physics SB RAS,
Irkutsk, Russia, 664033, Lermontova Str., 126a, POBox 291}
\ead{berng@iszf.irk.ru}

\begin{abstract}
In the paper within the approximation of the two-fluid magnetohydrodynamics
and geometrooptical approximation the dispersion relation was found
for ionacoustic instabilities of the ionospheric plasma at 80-200km
altitudes in three-dimensional weakly irregular ionosphere. Low freqeuncy
solution was found. The difference between obtained and standard solution
becomes significant at altitudes above 140 km. As the analysis shown
in this case the solution grows with time. The conditions for existence
of such solution are the presence of co-directed electron density gradients
and electron drifts and perpendicularity of line-of-sight to the magnetic
field. The necessary conditions regularly exist at the magnetic equator.
Detailed analysis has shown that this solution corresponds to well-known
150km equatorial echo and explains some of its statistical characteristics
observed experimentally. 

\end{abstract}

%Uncomment for PACS numbers title message
%\pacs{00.00, 20.00, 42.10}
\pacs{52.25.Xz, 52.30.Ex, 52.35.Qz, 94.20.dt, 94.20.wf}

% Keywords required only for MST, PB, PMB, PM, JOA, JOB? 
%\vspace{2pc}
%\noindent{\it Keywords}: Article preparation, IOP journals
% Uncomment for Submitted to journal title message
\submitto{\PPCF}
% Comment out if separate title page not required
\maketitle

\section{Introduction}

One of the important fields of modern ionospheric investigations is
the study of small-scale irregularities in E- and F- layers of the
ionosphere, that affect radiowave propagation and functionality of
different HF and UHF radiotools. One of the most investigated types
of irregularities is E- and F- layer irregularities produced as a
result of growth of two-stream and gradient-drift instabilities. The
theory of such instabilities is under development for a long time
but still is not finished \cite{Farley1963,Buneman1963,Kelley1989,Dimant1995}.

The usual condition for the growth of such irregularities is the requirement
of different velocities of electrons and ions, most significant at
altitudes 80-120km. At these heights ions are 'unmagnetized' - their
motion is controlled by neutral component motion. At the same time
the electrons motion is controlled by auxiliary electric and magnetic
fields - electrons are 'magnetized' \cite{Farley1963,Buneman1963,Kelley1989}.

But it is clear that this requirement significantly limits the validity
region of the current instabilities theories, that is why it is important
to obtain a theory of ionacoustic instabilities without this limitation
(see for example \cite{KaganKelley2000}).

\section{Basic equations}

\subsection{Two-fluid magnetohydrodynamic equations}

As basic equations for obtaining dispersion relation we will use two-fluid
magnetohydrodynamics (MHD) equations in form \cite{Galant1977}:

\begin{equation}
\left\{ \begin{array}{l}
\frac{dN_{\alpha}}{dt}+\overrightarrow{\bigtriangledown}(N_{\alpha}\overrightarrow{V_{\alpha}})=0\\
m_{\alpha}N_{\alpha}\frac{\partial\overrightarrow{V}_{\alpha}}{\partial t}=-Z_{\alpha}eN_{\alpha}(\overrightarrow{E}+\overrightarrow{V}_{\alpha}\times\overrightarrow{B})+\\
\,\,\,-m_{\alpha}N_{\alpha}(\overrightarrow{V}_{\alpha}\overrightarrow{\bigtriangledown})\overrightarrow{V}_{\alpha}-\overrightarrow{\bigtriangledown}(T_{\alpha}N_{\alpha})-N_{\alpha}m_{\alpha}\overrightarrow{V}_{\alpha}\nu_{\alpha n}^{t,\mu}\\
\overrightarrow{E}=\overrightarrow{E}_{0}-\overrightarrow{\bigtriangledown}\Phi\\
-\bigtriangledown^{2}\Phi=e(Z_{i}N_{i}+N_{e})\\
\overrightarrow{B}=\overrightarrow{B}_{0}\end{array}\right.\label{eq:2.1.1}\end{equation}
 where $\alpha=e,i$, normalized elastic collision frequencies:

\begin{equation}
\nu_{\alpha n}^{t,\mu}=\frac{\mu_{\alpha n}}{m_{\alpha}}\nu_{\alpha n}^{t}\label{eq:3.1.2}\end{equation}
 $\mu_{\alpha n}$ is effective mass of charged particles during elastic
collisions with neutrals (for electrons $\mu_{en}$ is very close
to electron mass, for ions $\mu_{in}$ could vary around half of ion
mass):

\begin{equation}
\mu_{\alpha n}=\frac{m_{\alpha}m_{n}}{m_{\alpha}+m_{n}}\label{eq:2.1.2a}\end{equation}
 and $\nu_{\alpha n}^{t}$ - elastic collision frequency of the ions
(electrons) with neutrals. We also suppose here that the charged particles
do not interact with each other through the collisions and interact
only through electromagnetic field. We also exclude all the viscidity
effects, that usually are not taken into account \cite{Galant1977}.
In this case we take into consideration (following to the \cite{Galant1977})
only elastic collisions. In detail the approximations used are listed
in Appendix A. Most of these approximations are valid at heights below
200km both for quiet and disturbed ionospheric conditions.

It must also be noted, that we throw out a lot of terms from the MHD
equations (\ref{eq:2.1.1}): ambient magnetic $\overrightarrow{B}_{0}$
and electric $\overrightarrow{E}_{0}$ fields are supposed to be constant;
any magnetic field variations are not taken into consideration; gravitation
field is not taken into consideration; recombination and ionization
processes are not taken into consideration; neutral component motion
is neglected. This allows us to neglect a lot of instabilities and
effects (see, for example \cite{Akhiezer1974}) and simplify the analysis.

\subsection{Zero order solution - quasihomogeneous and static case}

Zero order approximation connects nondisturbed (quasihomogeneous and
static) values of the particles density $N_{\alpha0}$, average motion
speed $\overrightarrow{V}_{\alpha0}$, ambient electrical $\overrightarrow{E}_{0}$
and magnetic $\overrightarrow{B}_{0}$ fields and collision frequencies.
When density, fields and collision frequencies are given, the zero
order approximation defines average motion speed of charged particles
in any point of space and time.

As one can see from (\ref{eq:2.1.1}), the zero-order approximation
is defined by the system:

\begin{equation}
\left\{ \begin{array}{l}
\overrightarrow{\bigtriangledown}\left(N_{\alpha0}\overrightarrow{V}_{\alpha0}\right)=0\\
0=Z_{\alpha}eN_{\alpha0}\overrightarrow{E}_{0}+Z_{\alpha}eN_{\alpha0}\overrightarrow{V}_{\alpha0}\times\overrightarrow{B}_{0}+\\
\,\,\,-m_{\alpha}N_{\alpha0}(\overrightarrow{V}_{\alpha0}\overrightarrow{\bigtriangledown})\overrightarrow{V}_{\alpha0}-\overrightarrow{\bigtriangledown}(T_{\alpha0}N_{\alpha0})+\\
\,\,\,-N_{\alpha0}m_{\alpha}\overrightarrow{V}_{\alpha0}\nu_{\alpha n}^{t,\mu}\\
0=e(Z_{i}N_{i0}+N_{e0})\end{array}\right.\label{eq:3.2.1}\end{equation}

In the simplest case of weak velocity gradients, when we could neglect
Lagrange term $(\overrightarrow{V}_{\alpha0}\overrightarrow{\bigtriangledown})\overrightarrow{V}_{\alpha0}$,
the system (\ref{eq:3.2.1}) has a well known solution \cite{Gurevich1973,Galant1977}:

\begin{equation}
\overrightarrow{V}_{\alpha0}=-\widehat{D_{\alpha}}\frac{\overrightarrow{\nabla}N_{\alpha}}{N_{\alpha0}}-\widehat{D_{T\alpha}}\frac{\overrightarrow{\nabla}T_{\alpha}}{T_{\alpha}}-\frac{\widehat{\sigma_{\alpha}}}{Z_{\alpha}eN_{\alpha0}}\overrightarrow{E}_{0}\label{eq:3.2.2}\end{equation}
 where operators of diffusion $\widehat{D_{\alpha}}$, thermodiffusion
$\widehat{D_{T\alpha}}$ and conductivity $\widehat{\sigma_{\alpha}}$
are:

\begin{equation}
\widehat{A_{\alpha}}=\left[\begin{array}{ccc}
A_{H\alpha} & -\Omega_{\alpha}A'_{H\alpha} & 0\\
\Omega_{\alpha}A'_{H\alpha} & A_{H\alpha} & 0\\
0 & 0 & A_{\alpha}\end{array}\right]\label{eq:3.2.3}\end{equation}

\begin{equation}
\widehat{A_{T\alpha}}=\left[\begin{array}{ccc}
A_{T\alpha(1,1)} & -A_{T\alpha(1,2)} & 0\\
A_{T\alpha(1,2)} & A_{T\alpha(1,1)} & 0\\
0 & 0 & \left(A_{\alpha}+\frac{T_{\alpha}dA_{\alpha}}{dT_{\alpha}}\right)\end{array}\right]\label{eq:3.2.4}\end{equation}

\begin{equation}
A_{T\alpha(1,1)}=\left(A_{H\alpha}+\frac{T_{\alpha}dA_{H\alpha}}{dT_{\alpha}}\right)\label{eq:3.2.4b}\end{equation}

\begin{equation}
A_{T\alpha(1,2)}=\Omega_{\alpha}\left(A'_{H\alpha}+\frac{T_{\alpha}dA'_{H\alpha}}{dT_{\alpha}}\right)\label{eq:3.2.4c}\end{equation}

\begin{equation}
\widehat{D_{\alpha}}=\frac{T_{\alpha}}{m_{\alpha}}\widehat{A_{\alpha}}\label{eq:3.2.5}\end{equation}

\begin{equation}
\widehat{D_{T\alpha}}=\frac{T_{\alpha}}{m_{\alpha}}\widehat{A_{T\alpha}}\label{eq:3.2.6}\end{equation}

\begin{equation}
\widehat{\sigma_{\alpha}}=\frac{Z^{2}e^{2}N_{\alpha}}{m_{\alpha}}\widehat{A_{\alpha}}\label{eq:3.2.7}\end{equation}
 \begin{eqnarray}
A_{H\alpha} & = & \frac{\nu_{\alpha}K_{\sigma}(\frac{\Omega_{\alpha}}{\nu_{\alpha}})}{\Omega_{\alpha}^{2}+\nu_{\alpha}^{2}}\nonumber \\
A'_{H\alpha} & = & \frac{K_{\varepsilon}(\frac{\Omega_{\alpha}}{\nu_{\alpha}})}{\Omega_{\alpha}^{2}+\nu_{\alpha}^{2}}\nonumber \\
A_{\alpha} & = & \frac{K_{\sigma}(0)}{\nu_{\alpha}}\label{eq:3.2.8}\end{eqnarray}
 and functions $K_{\sigma}(x),K_{\varepsilon}(x)$ are tabulated (for
example in \cite{Gurevich1973}) for taking into account not only
MHD effects, but kinetic effects too.

It is important to note that for our next consideration the exact
expression (\ref{eq:3.2.2}) for zero-order solution is not very significant
for us, and below we only suggest that the solution $\overrightarrow{V}_{\alpha0}=\overrightarrow{V}_{\alpha0}\left(N_{\alpha},T_{\alpha},\overrightarrow{E}_{0},\overrightarrow{B}_{0},\nu_{\alpha n}^{t,\mu}\right)$
exists and is unambiguously determined by its arguments. So by the
zero-order solution we mean an equation (\ref{eq:3.2.1}) that defines
an average motion speed $\overrightarrow{V}_{\alpha0}$ as a function
of ambient conditions and which could be solved analytically (\ref{eq:3.2.2}-\ref{eq:3.2.8})
in simple cases or numerically in more complex cases.

\subsection{First order solution - nonstatic inhomogeneous case}

One of standard approaches to the MHD equations analysis is a geometrooptical
(GO) approximation, the validity of which is defined by smallness
of the parameter

\begin{equation}
\mu=\left(\overrightarrow{k}\left(\frac{\overrightarrow{\nabla}P}{P}\right)\right)^{-1}<<1\label{eq:2.2.1}\end{equation}
 where $\overrightarrow{k}$ is irregularities wave vector and $\frac{\overrightarrow{\nabla}P}{P}$
typical range of changes of parameter $P$ (for example electron density).

When the GO approximation is valid, the solution for small variations
of the parameters $N,\overrightarrow{V},\overrightarrow{E}$ can be
found in form:

\begin{equation}
\begin{array}{l}
\delta N_{\alpha}(\overrightarrow{r},t)=e^{-i\psi(\mu\overrightarrow{r},\mu t)/\mu}N_{\alpha1}(\mu\overrightarrow{r},\mu t)\\
\delta\overrightarrow{V}_{\alpha}(\overrightarrow{r},t)=e^{-i\psi(\mu\overrightarrow{r},\mu t)/\mu}\overrightarrow{V}_{\alpha1}(\mu\overrightarrow{r},\mu t)\\
\delta\overrightarrow{E}(\overrightarrow{r},t)=-\overrightarrow{\bigtriangledown}\left(e^{-i\psi(\mu\overrightarrow{r},\mu t)/\mu}\Phi_{1}(\mu\overrightarrow{r},\mu t)\right)\end{array}\label{eq:2.2.3}\end{equation}

Geometrooptical phase $\psi(\mu\overrightarrow{r},\mu t)/\mu$ (or
eikonal) for plane waves is related to wave vector $\overrightarrow{k}$
and complex frequency of the wave $\omega+i\gamma$ by the following
definitions:

\begin{eqnarray}
\overrightarrow{k} & = & -\overrightarrow{\bigtriangledown}\psi(\mu\overrightarrow{r},\mu t)/\mu\nonumber \\
\omega+i\gamma & = & \frac{\partial\psi(\mu\overrightarrow{r},\mu t)}{\mu\partial t}\label{eq:2.2.4}\end{eqnarray}

The first approximation gave us the system of equations:

\begin{equation}
\left\{ \begin{array}{l}
P_{\alpha1}N_{\alpha1}+\overrightarrow{P}_{\alpha2}\overrightarrow{V}_{\alpha1}=0\\
\overrightarrow{P_{\alpha3}}N_{\alpha1}+\widehat{P_{\alpha0}}\overrightarrow{V}_{\alpha1}+\overrightarrow{P_{\alpha4}}\Phi_{1}=0\\
Z_{i}N_{i1}=-N_{e1}-\frac{1}{e}\Phi_{1}(\overrightarrow{\bigtriangledown}\psi)^{2}\end{array}\right.\label{eq:3.3.1}\end{equation}

where, by taking into account the zero-order approximation (\ref{eq:3.2.1}):

\begin{equation}
P_{\alpha1}=\left(-\frac{\overrightarrow{V}_{\alpha0}\overrightarrow{\bigtriangledown}N_{\alpha0}}{N_{\alpha0}}+i\overrightarrow{V}_{\alpha0}\overrightarrow{\bigtriangledown}\psi+i\frac{\partial\psi}{\partial t}\right)\label{eq:3.3.3}\end{equation}

\begin{equation}
P_{\alpha2}=\left(\overrightarrow{\bigtriangledown}N_{\alpha0}+iN_{\alpha0}\overrightarrow{\bigtriangledown}\psi\right)\label{eq:3.3.4}\end{equation}

\begin{equation}
\overrightarrow{P_{\alpha3}}=T_{\alpha}\left(-\left(i\overrightarrow{\bigtriangledown}\psi\right)+\left(\frac{\overrightarrow{\bigtriangledown}N_{\alpha0}}{N_{\alpha0}}\right)\right)\label{eq:3.3.5}\end{equation}

\begin{equation}
P_{\alpha4}=iZ_{\alpha}eN_{\alpha0}\left(\overrightarrow{\bigtriangledown}\psi\right)\label{eq:3.3.6}\end{equation}

\begin{equation}
\widehat{P_{\alpha0}}\overrightarrow{V}_{\alpha1}=P_{\alpha5}\overrightarrow{V}_{\alpha1}+\overrightarrow{V}_{\alpha1}\times\overrightarrow{P_{\alpha6}}+\widehat{P_{\alpha7}}\overrightarrow{V}_{\alpha1}\label{eq:3.3.7}\end{equation}

\begin{equation}
P_{\alpha5}=-m_{\alpha}N_{\alpha0}\left(i\left(\frac{\partial\psi}{\partial t}\right)+i\left(\overrightarrow{V}_{\alpha0}\overrightarrow{\bigtriangledown}\psi\right)+\nu_{\alpha n}^{t,\mu}\right)\label{eq:3.3.8}\end{equation}

\begin{equation}
\overrightarrow{P}_{\alpha6}=-Z_{\alpha}eN_{\alpha0}\overrightarrow{B}_{0}\label{eq:3.3.9}\end{equation}

\begin{equation}
\widehat{P_{\alpha7}}\overrightarrow{V}_{\alpha1}=-m_{\alpha}\left(\left(N_{\alpha0}\overrightarrow{V}_{\alpha1}\overrightarrow{\bigtriangledown}\right)\overrightarrow{V}_{\alpha0}\right)\label{eq:3.3.10}\end{equation}

To make the following analysis easier, the system (\ref{eq:3.3.1})
is written in operator form, where operators $P_{n}$ are matrix operators
in partial derivatives over the eikonal $\psi$. It is clear that
in this form the system looks pretty simple and solvable.

\section{Dispersion relation}

\subsection{Obtaining the dispersion relation}

From (\ref{eq:3.3.1}) one can see that the system is linear and,
in case of existence and uniqueness of the inverse operator $\widehat{P}_{\alpha0}^{-1}$
(\ref{eq:3.3.7}) it can be solved. After excluding $\overrightarrow{V}_{\alpha1}$
from (\ref{eq:3.3.1}) the equation connecting the density $N_{\alpha1}$
and electric potential $\Phi_{1}$ variations has the following form:

\begin{equation}
C_{1\alpha}(\psi)N_{\alpha1}+C_{2\alpha}(\psi)\Phi_{1}=0\label{eq:3.4.1}\end{equation}
 where coefficients are:

\begin{equation}
\begin{array}{l}
C_{1\alpha}(\psi)=A(...)\left(P_{\alpha1}-\overrightarrow{P}_{\alpha2}\left(\widehat{P}_{\alpha0}^{-1}\overrightarrow{P}_{\alpha3}\right)\right)\\
C_{2\alpha}(\psi)=-A(...)\overrightarrow{P}_{\alpha2}\left(\widehat{P}_{\alpha0}^{-1}\overrightarrow{P}_{\alpha4}\right)\end{array}\label{eq:3.4.2}\end{equation}
 and $A(...)$ - an arbitrary function of arbitrary parameters that
does not have zero values at the investigated region.

Now we can recall that our plasma has two types of particles and its
characteristics are defined by the system of equations:

\begin{equation}
\left\{ \begin{array}{l}
C_{1i}(\psi)N_{i1}+C_{2i}(\psi)\Phi_{1}=0\\
C_{1e}(\psi)N_{e1}+C_{2e}(\psi)\Phi_{1}=0\\
Z_{i}N_{i1}=-N_{e1}-\frac{1}{e}\Phi_{1}(\overrightarrow{\bigtriangledown}\psi)^{2}\end{array}\right.\label{eq:3.4.3}\end{equation}

Sometimes, for example, when analyzing thermal variations of electron
density (that cause incoherent scattering), the self-coordinated term
$\frac{1}{e}\Phi_{1}(\overrightarrow{\bigtriangledown}\psi)^{2}$
in (\ref{eq:3.4.3}) can not be neglected - scatterers size has order
of Debye length and this term becomes significant. But in this very
task we can neglect this term, following to many authors (see for
example \cite{Dimant1995}). 

It is clear that existence of solution of (\ref{eq:3.4.3}) is determined
by consistency of these equations. The consistency condition in our
case has the following form:

\begin{equation}
C_{1i}(\psi)C_{2e}(\psi)+Z_{i}C_{1e}(\psi)C_{2i}(\psi)=0\label{eq:3.4.4}\end{equation}

It connects different partial derivatives over the eikonal $\psi$
with each other and can be referred as dispersion relation. As one
can see, the dispersion relation has symmetrical (as it was expected
earlier) form.

\subsection{IAQV approximation}

It is clear that existence of dispersion relation and its exact form
(\ref{eq:3.4.4}) depend on existence and properties of inverse operator
$\widehat{P}_{\alpha0}^{-1}$. As it has been shown in Appendix B,
the inverse operator can be easily found in case when Lagrange term
$\widehat{P_{7}}$ in $\widehat{P}_{\alpha0}$ can be neglected. Below
we call this approximation as 'Irregularities under approximation
of quasihomogeneous velocity' (IAQV). As preliminary analysis has
shown, this approximation is valid for wavenumbers 0.1-10 $m^{-1}$
under most ionospheric conditions at altitudes below 200km and for
variations of average parameters not faster than 100m (for faster
changes the GO approximation becomes incorrect).

In the IAQV approximation the inverse operator $\widehat{P}_{\alpha0}$
has the following simple form:

\begin{equation}
\widehat{P}_{\alpha0}^{-1}\overrightarrow{f}=\frac{\widehat{b}\left(\widehat{b}\overrightarrow{f}\right)P_{\alpha6}^{2}+P_{\alpha5}^{2}\overrightarrow{f}+P_{\alpha5}P_{\alpha6}(\widehat{b}\times\overrightarrow{f})}{P_{\alpha5}\left(P_{\alpha6}^{2}+P_{\alpha5}^{2}\right)}\label{eq:3.5.1}\end{equation}
 where $\widehat{b}$ - unity vector in direction of $\overrightarrow{P}_{\alpha6}$
(and antiparallel to the magnetic field).

It must be noted that IAQV approximation does not mean neglecting
the Lagrange term $\left(\overrightarrow{V}_{\alpha}\overrightarrow{\bigtriangledown}\right)\overrightarrow{V}_{\alpha}$
in basic equations (\ref{eq:2.1.1}), but only neglecting $\left(\overrightarrow{V}_{\alpha1}\overrightarrow{\bigtriangledown}\right)\overrightarrow{V}_{\alpha0}$
term in operator $\widehat{P}_{\alpha0}$ (\ref{eq:3.3.7}), all the
other terms are the same.

Summarizing, the dispersion relation in IAQV approximation has the
following form (\ref{eq:3.4.4}, \ref{eq:3.3.3}-\ref{eq:3.3.9},
\ref{eq:3.5.1}).

\subsection{The basic structure of the dispersion relation}

Let us briefly analyze the structure of the dispersion relation by
defining function that does not have zeroes:

\begin{equation}
A(...)=P_{\alpha5}\left(P_{\alpha6}^{2}+P_{\alpha5}^{2}\right)\label{eq:4.1.1}\end{equation}

This leads to the following coefficients of the dispersion relation
(\ref{eq:3.4.4}):

\begin{equation}
\begin{array}{l}
C_{1\alpha}(\psi)=P_{\alpha1}P_{\alpha5}\left(P_{\alpha6}^{2}+P_{\alpha5}^{2}\right)+\\
\,\,\,-\left(\overrightarrow{P}_{\alpha2}\widehat{b}\right)\left(\widehat{b}\overrightarrow{P}_{\alpha3}\right)P_{\alpha6}^{2}-P_{\alpha5}^{2}\overrightarrow{P}_{\alpha2}\overrightarrow{P}_{\alpha3}+\\
\,\,\,-P_{\alpha5}P_{\alpha6}\overrightarrow{P}_{\alpha2}(\widehat{b}\times\overrightarrow{P}_{\alpha3})\\
C_{2\alpha}(\psi)=-\left(\widehat{b}\overrightarrow{P}_{\alpha2}\right)\left(\widehat{b}\overrightarrow{P}_{\alpha4}\right)P_{\alpha6}^{2}-P_{\alpha5}^{2}\overrightarrow{P}_{\alpha2}\overrightarrow{P}_{\alpha4}+\\
\,\,\,-P_{\alpha5}P_{\alpha6}\overrightarrow{P}_{\alpha2}(\widehat{b}\times\overrightarrow{P}_{\alpha4})\end{array}\label{eq:4.1.2}\end{equation}

When taking into account (\ref{eq:3.3.3}-\ref{eq:3.3.9}) it becomes
clear that coefficients (\ref{eq:4.1.2}) are polynomials over the
$\left(\frac{\partial\psi}{\partial t}\right)$ and have the form:

\begin{equation}
\begin{array}{l}
C_{1\alpha}(\psi)=R_{1\alpha4}\left(\frac{\partial\psi}{\partial t}\right)^{4}+R_{1\alpha3}\left(\frac{\partial\psi}{\partial t}\right)^{3}+\\
\,\,\,\,\,\,\,\,\,\,\,\,+R_{1\alpha2}\left(\frac{\partial\psi}{\partial t}\right)^{2}+R_{1\alpha1}\left(\frac{\partial\psi}{\partial t}\right)+R_{1\alpha0}\\
C_{2\alpha}(\psi)=R_{2\alpha2}\left(\frac{\partial\psi}{\partial t}\right)^{2}+R_{2\alpha1}\left(\frac{\partial\psi}{\partial t}\right)+R_{2\alpha0}\end{array}\label{eq:4.1.3}\end{equation}

From this consideration it becomes clear that dispersion relation
(\ref{eq:3.4.4}) is a 6th order polynomial over the $\left(\frac{\partial\psi}{\partial t}\right)$
and has no more than 6 solutions.

\subsection{Simple representation of coefficients}

To simplify the solution technique in homogeneous case, in the work
\cite{FejerFarleyEtAl1975} a new complex variable was defined:

\begin{equation}
\widetilde{\omega}_{\alpha}=\frac{\partial\psi}{\partial t}+\overrightarrow{V}_{\alpha0}\overrightarrow{\bigtriangledown}\psi\label{eq:4.2.1}\end{equation}

In our inhomogeneous case we define the following new variables:

\begin{equation}
\overrightarrow{K}_{N}=\frac{\overrightarrow{\bigtriangledown}N_{\alpha0}}{N_{\alpha0}}\label{eq:4.2.2}\end{equation}

\begin{equation}
\widetilde{\omega}_{\alpha N}=\frac{\partial\psi}{\partial t}+\overrightarrow{V}_{\alpha0}\overrightarrow{\bigtriangledown}\psi+i\overrightarrow{K}_{N}\overrightarrow{V}_{\alpha0}\label{eq:4.2.3}\end{equation}

\begin{equation}
\overrightarrow{k}_{N}=\overrightarrow{\bigtriangledown}\psi+i\overrightarrow{K}_{N}\label{eq:4.2.4}\end{equation}

\begin{equation}
\nu_{\alpha nN}^{t,\mu}=\nu_{\alpha n}^{t,\mu}+\overrightarrow{K}_{N}\overrightarrow{V}_{\alpha0}\label{eq:4.2.5}\end{equation}

In this case the operators (\ref{eq:3.3.3}-\ref{eq:3.3.9}) become:

\begin{equation}
P_{\alpha1}=i\left(\widetilde{\omega}_{\alpha N}\right)\label{eq:4.2.6}\end{equation}

\begin{equation}
\overrightarrow{P_{\alpha2}}=iN_{\alpha0}\overrightarrow{k}_{N}^{*}\label{eq:4.2.7}\end{equation}

\begin{equation}
\overrightarrow{P_{\alpha3}}=-iT_{\alpha}\overrightarrow{k}_{N}\label{eq:4.2.8}\end{equation}

\begin{equation}
\overrightarrow{P_{\alpha4}}=iZ_{\alpha}eN_{\alpha0}\left(\frac{\overrightarrow{k}_{N}+\overrightarrow{k}_{N}^{*}}{2}\right)\label{eq:4.2.9}\end{equation}

\begin{equation}
P_{\alpha5}=\left(-im_{\alpha}N_{\alpha0}\right)\left(\widetilde{\omega}_{\alpha N}-i\nu_{\alpha nN}^{t,\mu}\right)\label{eq:4.2.10}\end{equation}

\begin{equation}
\overrightarrow{P}_{\alpha6}=-Z_{\alpha}eN_{\alpha0}\overrightarrow{B}_{0}\label{eq:4.2.11}\end{equation}

\begin{equation}
P_{\alpha6}=Z_{\alpha}eN_{\alpha0}B_{0}\label{eq:4.2.12}\end{equation}

where {*} is a complex conjugation.

Summarizing, the dispersion relation has the form (\ref{eq:3.4.4}),
where its coefficients are defined by (\ref{eq:4.1.2},\ref{eq:4.2.2}-\ref{eq:4.2.12})

\section{Dispersion relation for weak gradients case}

\subsection{Weak gradients approximation}

The vectors $\overrightarrow{P_{\alpha2}},\overrightarrow{P_{\alpha3}},\overrightarrow{P_{\alpha4}}$
in dispersion relation are parallel in case of homogeneous ionosphere
and not parallel in case of inhomogeneous ionosphere. After substituting
$\overrightarrow{P_{\alpha2}},\overrightarrow{P_{\alpha3}},\overrightarrow{P_{\alpha4}}$
(\ref{eq:4.2.7}-\ref{eq:4.2.9}) into (\ref{eq:4.1.2}) and taking
into account the properties of vector product we obtain the following:

\begin{equation}
\begin{array}{l}
C_{1\alpha}(\psi)=P_{\alpha1}P_{\alpha5}\left(P_{\alpha6}^{2}+P_{\alpha5}^{2}\right)+\\
\,\,\,-N_{\alpha0}T_{\alpha}\left(\left|\overrightarrow{k}_{N}\widehat{b}\right|^{2}P_{\alpha6}^{2}+\left|\overrightarrow{k}_{N}\right|^{2}P_{\alpha5}^{2}\right)+\\
\,\,\,+i2N_{\alpha0}T_{\alpha}P_{\alpha5}P_{\alpha6}Im\left(\overrightarrow{k}_{N}\right)(\widehat{b}\times Re(\overrightarrow{k}_{N}))\\
C_{2\alpha}(\psi)=Z_{\alpha}eN_{\alpha0}^{2}\left(P_{\alpha6}^{2}\left(\widehat{b}Re(\overrightarrow{k}_{N})\right)^{2}\right)+\\
\,\,\,+Z_{\alpha}eN_{\alpha0}^{2}\left(P_{\alpha5}^{2}\left(Re(\overrightarrow{k}_{N})\right)^{2}\right)+\\
\,\,\,-iZ_{\alpha}eN_{\alpha0}^{2}\left(\left(Re(\overrightarrow{k}_{N})\right)\left(Im(\overrightarrow{k}_{N})\right)P_{\alpha5}^{2}\right)+\\
\,\,\,-iZ_{\alpha}eN_{\alpha0}^{2}\left(P_{\alpha5}P_{\alpha6}Im\left(\overrightarrow{k}_{N}\right)\left(\widehat{b}\times Re(\overrightarrow{k}_{N})\right)\right)+\\
\,\,\,-iZ_{\alpha}eN_{\alpha0}^{2}\left(\left(\widehat{b}Re(\overrightarrow{k}_{N})\right)\left(\widehat{b}Im(\overrightarrow{k}_{N})\right)P_{\alpha6}^{2}\right)\end{array}\label{eq:4.2.17}\end{equation}

The first two terms in each relation (\ref{eq:4.2.17}) are the terms
that correspond both to the inhomogeneous and homogeneous dispersion
relations, the last terms correspond only to changes of dispersion
relation due to inhomogeneities presence. It is clear that in both
cases (gradients are parallel to the magnetic field and gradients
are perpendicular to the magnetic field) the difference between dispersion
relation for homogeneous case and for inhomogeneous one does exist.
But if the gradients are weak enough (or wavenumbers are high enough)
we can neglect the changes of dispersion relation and solve only simplified
one, that corresponds to the homogeneous case:

\begin{equation}
\begin{array}{l}
C_{1\alpha}(\psi)=P_{\alpha1}P_{\alpha5}\left(P_{\alpha6}^{2}+P_{\alpha5}^{2}\right)+\\
\,\,\,-N_{\alpha0}T_{\alpha}\left(\left|\overrightarrow{k}_{N}\widehat{b}\right|^{2}P_{\alpha6}^{2}+\left|\overrightarrow{k}_{N}\right|^{2}P_{\alpha5}^{2}\right)\\
C_{2\alpha}(\psi)=Z_{\alpha}eN_{\alpha0}^{2}\left(P_{\alpha6}^{2}\left(\widehat{b}Re(\overrightarrow{k}_{N})\right)^{2}+\right.\\
\,\,\,\,\,\,\,\,\,\,\,\,\,\,\,\,\,\,\,\,\,\,\,\,\left.+P_{\alpha5}^{2}\left(Re(\overrightarrow{k}_{N})\right)^{2}\right)\end{array}\label{eq:4.2.18}\end{equation}

As one can see, this approximation is valid, when:

\begin{equation}
\frac{\left(Re(\overrightarrow{k}_{N})\right)}{\left(Im(\overrightarrow{k}_{N})\right)}>>1\label{eq:4.2.18b}\end{equation}

\begin{equation}
\frac{\left|P_{\alpha5}\right|^{2}\left(Re(\overrightarrow{k}_{N})\right)^{2}}{\left(\widehat{b}Re(\overrightarrow{k}_{N})\right)\left(\widehat{b}Im(\overrightarrow{k}_{N})\right)P_{\alpha6}^{2}}>>1\label{eq:4.2.18c}\end{equation}

\begin{equation}
\frac{\left|P_{\alpha5}\right|\left(Re(\overrightarrow{k}_{N})\right)^{2}}{\left|P_{\alpha6}\right|Im\left(\overrightarrow{k}_{N}\right)\left(\widehat{b}\times Re(\overrightarrow{k}_{N})\right)}>>1\label{eq:4.2.18d}\end{equation}

The condition (\ref{eq:4.2.18b}) is equivalent to the GO validity
condition (\ref{eq:2.2.1}):

\[
\left|\overrightarrow{k}_{N}\right|>>\left|\overrightarrow{K}_{N}\right|\]

The condition (\ref{eq:4.2.18c}) is valid when analyzing scattering
almost perpendicular to the magnetic field or when gradients are sufficiently
small:

\[
\left|\overrightarrow{k}_{N}\right|\frac{\left|P_{\alpha5}\right|^{2}}{\left|P_{\alpha6}\right|^{2}}>>\frac{\left(\widehat{b}Re(\overrightarrow{k}_{N})\right)}{\left(Re(\overrightarrow{k}_{N})\right)}\left(\widehat{b}\overrightarrow{K}_{N}\right)\]

And the last condition (\ref{eq:4.2.18d}) is valid when gradients
are small enough:

\begin{equation}
\left|\overrightarrow{k}_{N}\right|\frac{\left|P_{\alpha5}\right|}{\left|P_{\alpha6}\right|}>>\left|\overrightarrow{K}_{N}\right|\label{eq:4.2.18e}\end{equation}

So the condition (\ref{eq:4.2.18b}) is always valid, condition (\ref{eq:4.2.18c})
is valid when we investigate the instabilities near the perpendicular
to the magnetic field and the condition (\ref{eq:4.2.18e}) becomes
only critical limitation for the approximation (\ref{eq:4.2.18})
of initial formula (\ref{eq:4.2.17}).

\subsection{Ionospheric parameters}

To create a correct dispersion relation for heights 80-200km we should
choose the correct approximations for ionospheric plasma. The most
important plasma parameters are thermal velocities, hyrofrequencies
and frequencies of collisions with neutrals. These approximate parameters
are shown at the Table 1 (calculated for mid-latitude ionosphere using
models MSIS, IGRF and IRI)

We also suggest that wavenumbers are within 0.1-10m (sounding frequencies
15-1500MHz), and drift velocities $\overrightarrow{V}_{\alpha0}$
do not exceed 3000m/s.

It is clear, that at altitudes 80-200km the following approximations
are valid: $\sqrt{\frac{T_{e}}{m_{e}}}\left(\overrightarrow{\bigtriangledown}\psi\right)<<\Omega_{e}$
- electron hyrofrequency much higher than their thermal speed; $\left(\overrightarrow{V}_{e0}\overrightarrow{\bigtriangledown}\psi\right)<<\Omega_{e}$
- electron hyrofrequency much higher than their average speed (even
in disturbed conditions); $\left(\overrightarrow{V}_{e0}\right)<<\sqrt{\frac{T_{e}}{m_{e}}}$
- electron thermal speed much higher than their average speed; $\nu_{en}^{t,\mu}>>\nu_{in}^{t,\mu}$
electron-neutral collision frequency is much higher than ion-neutral
one; We also use weakly inhomogeneous ionosphere approximation: $\left|\overrightarrow{\bigtriangledown}\psi\right|>>\left|\frac{\overrightarrow{\bigtriangledown}N_{e0}}{N_{e0}}\right|$

Below there is a list of traditional approximations that are valid
for E-layer \cite{Farley1963,Buneman1963,Kelley1989}, but is not
valid for the whole region 80-200km: $\left(\overrightarrow{V}_{\alpha0}\overrightarrow{\bigtriangledown}\psi\right)<<\nu_{\alpha n}^{t,\mu}$
- is not valid for ions at altitudes above 120km, not valid for electrons
at heights above 180km under very disturbed conditions; $\left(\overrightarrow{V}_{i0}\overrightarrow{\bigtriangledown}\psi\right)<<\Omega_{i}$
- is not valid for disturbed conditions; $\Omega_{i}<<\nu_{in}^{t,\mu}$
- is not valid above 120km; $\Omega_{e}>>\nu_{en}^{t,\mu}$ - is not
valid at and below 80km. It also must be noted that during high disturbances
the effective ion-neutral collision frequency $\nu_{inN}^{t,\mu}$
, that is used in dispersion relation, becomes dependent on electron
density gradient and average ions velocity (\ref{eq:4.2.5}) and might
be increased (or decreased) depending on the ion motion direction
and gradients. It should be also noted that for high velocities the
effective ion-neutral collision frequency $\nu_{inN}^{t,\mu}$ at
heights approximately above 140-160km can become zero or negative.
So, in this case the traditional approximation \cite{Dimant1995}
of low Doppler drifts $|\widetilde{\omega}_{\alpha}|<<|\nu_{\alpha nN}^{t,\mu}|$
is also invalid.

\subsection{Dispersion relation for weak gradients}

By substituting (\ref{eq:4.2.6},\ref{eq:4.2.10},\ref{eq:4.2.11})
into (\ref{eq:4.2.18}) and after neglecting same non-zero multipliers:

\begin{equation}
\begin{array}{l}
C_{1\alpha}\left(\widetilde{\omega}_{\alpha N}\right)=\left(\widetilde{\omega}_{\alpha N}\right)\left(\widetilde{\omega}_{\alpha N}-i\nu_{\alpha nN}^{t,\mu}\right)m_{\alpha}\cdot\\
\cdot\left(\Omega_{\alpha}^{2}+\left(\nu_{\alpha nN}^{t,\mu}+i\widetilde{\omega}_{\alpha N}\right)^{2}\right)-\\
-T_{\alpha}\left(\left|\overrightarrow{k}_{N}\widehat{b}\right|^{2}\Omega_{\alpha}^{2}+\left|\overrightarrow{k}_{N}\right|^{2}\left(\nu_{\alpha nN}^{t,\mu}+i\widetilde{\omega}_{\alpha N}\right)^{2}\right)\\
C_{2\alpha}\left(\widetilde{\omega}_{\alpha N}\right)=Z_{\alpha}eN_{\alpha0}\cdot\\
\cdot\left(\Omega_{\alpha}^{2}\left(\widehat{b}Re(\overrightarrow{k}_{N})\right)^{2}+\left(\nu_{\alpha nN}^{t,\mu}+i\widetilde{\omega}_{\alpha N}\right)^{2}\left(Re(\overrightarrow{k}_{N})\right)^{2}\right)\end{array}\label{eq:4.3.2}\end{equation}

where

\begin{equation}
\Omega_{\alpha}=\frac{Z_{\alpha}eB_{0}}{m_{\alpha}}\label{eq:4.3.3}\end{equation}

is hyrofrequency.

Let us define new index $\beta$ to describe another charged component:
$\beta=e,i;\beta\neq\alpha$.

After defining the new parameters

\begin{equation}
\delta\widetilde{\omega}_{\beta\alpha N}=\widetilde{\omega}_{\beta N}-\widetilde{\omega}_{\alpha N}=\left(\overrightarrow{V}_{\alpha0}-\overrightarrow{V}_{\beta0}\right)\left(\overrightarrow{k}-i\overrightarrow{K}_{N}\right)\label{eq:4.3.4}\end{equation}

\begin{equation}
\overrightarrow{k}=-Re(\overrightarrow{k}_{N})=-\overrightarrow{\bigtriangledown}\psi\label{eq:4.3.4b}\end{equation}

we will obtain the relations for second charged component as a function
of the same parameter $\widetilde{\omega}_{\alpha N}$:

\begin{equation}
\begin{array}{l}
C_{1\beta}\left(\widetilde{\omega}_{\alpha N}\right)=\left(\widetilde{\omega}_{\alpha N}+\delta\widetilde{\omega}_{\beta\alpha N}\right)\left(\widetilde{\omega}_{\alpha N}+\delta\widetilde{\omega}_{\beta\alpha N}-i\nu_{\beta nN}^{t,\mu}\right)\cdot\\
\cdot m_{\beta}\left(\Omega_{\beta}^{2}+\left(\nu_{\beta nN}^{t,\mu}+i\widetilde{\omega}_{\alpha N}+i\delta\widetilde{\omega}_{\beta\alpha N}\right)^{2}\right)-\\
-T_{\beta}\left(\left|\overrightarrow{k}\widehat{b}\right|^{2}\Omega_{\beta}^{2}+\left|\overrightarrow{k}\right|^{2}\left(\nu_{\beta nN}^{t,\mu}+i\widetilde{\omega}_{\alpha N}+i\delta\widetilde{\omega}_{\beta\alpha N}\right)^{2}\right)\\
C_{2\beta}\left(\widetilde{\omega}_{\alpha N}\right)=Z_{\beta}eN_{\beta0}\left(\Omega_{\beta}^{2}\left(\widehat{b}\overrightarrow{k}\right)^{2}+\right.\\
\left.+\left(\nu_{\beta nN}^{t,\mu}+i\widetilde{\omega}_{\alpha N}+i\delta\widetilde{\omega}_{\beta\alpha N}\right)^{2}\left(\overrightarrow{k}\right)^{2}\right)\end{array}\label{eq:4.3.5}\end{equation}

Below we will analyze the dispersion relation (\ref{eq:3.4.4}) in
form:

\begin{equation}
f(x)=0\label{eq:4.3.6a}\end{equation}

where

\begin{equation}
f(x=\widetilde{\omega}_{\alpha N})=Z_{\beta}C_{1\alpha}(x)C_{2\beta}(x)+Z_{\alpha}C_{1\beta}(x)C_{2\alpha}(x)\label{eq:4.3.6}\end{equation}

By substituting (\ref{eq:4.3.2},\ref{eq:4.3.5}) into (\ref{eq:4.3.6}),
after neglecting non-zero multiplier, for single-charged ions ($Z_{i}=-1$)
(most frequent approximation in this region of altitudes) the dispersion
relation becomes the final one:

\begin{equation}
\begin{array}{l}
f(x)=m_{\alpha}x\left(x-i\nu_{\alpha nN}^{t,\mu}\right)\left(\Omega_{\alpha}^{2}+\left(\nu_{\alpha nN}^{t,\mu}+ix\right)^{2}\right)\cdot\\
\cdot\left(\Omega_{\beta}^{2}\left(\widehat{b}\overrightarrow{k}\right)^{2}+\left(\nu_{\beta nN}^{t,\mu}+i\left(x+\delta\widetilde{\omega}_{\beta\alpha N}\right)\right)^{2}\left(\overrightarrow{k}\right)^{2}\right)+\\
+m_{\beta}\left(x+\delta\widetilde{\omega}_{\beta\alpha N}\right)\left(x+\delta\widetilde{\omega}_{\beta\alpha N}-i\nu_{\beta nN}^{t,\mu}\right)\cdot\\
\cdot\left(\Omega_{\beta}^{2}+\left(\nu_{\beta nN}^{t,\mu}+i\left(x+\delta\widetilde{\omega}_{\beta\alpha N}\right)\right)^{2}\right)\cdot\\
\cdot\left(\Omega_{\alpha}^{2}\left(\widehat{b}\overrightarrow{k}\right)^{2}+\left(\nu_{\alpha nN}^{t,\mu}+ix\right)^{2}\left(\overrightarrow{k}\right)^{2}\right)+\\
-\left(T_{\alpha}+T_{\beta}\right)\left(\left|\widehat{b}\overrightarrow{k}\right|^{2}\Omega_{\alpha}^{2}+\left|\overrightarrow{k}\right|^{2}\left(\nu_{\alpha nN}^{t,\mu}+ix\right)^{2}\right)\cdot\\
\cdot\left(\Omega_{\beta}^{2}\left(\widehat{b}\overrightarrow{k}\right)^{2}+\left(\nu_{\beta nN}^{t,\mu}+i\left(x+\delta\widetilde{\omega}_{\beta\alpha N}\right)\right)^{2}\left(\overrightarrow{k}\right)^{2}\right)\end{array}\label{eq:4.3.7}\end{equation}
 where

\begin{equation}
x=\frac{\partial\psi}{\partial t}-\overrightarrow{V}_{\alpha0}\overrightarrow{k}+i\overrightarrow{K}_{N}\overrightarrow{V}_{\alpha0}\label{eq:4.3.8}\end{equation}
 and other parameters are defined by (\ref{eq:4.2.2}-\ref{eq:4.2.5},\ref{eq:4.3.4},\ref{eq:4.3.4b})
and by solution of the zero-order approximation (\ref{eq:3.2.1}).

From obtained solution $x_{0}$ of dispersion relation (\ref{eq:4.3.7})
for given altitude dependence of the parameters one can always obtain
the actual irregularity frequencies and decrements using relation:

\begin{equation}
\frac{\partial\psi}{\partial t}=x_{0}+\overrightarrow{V}_{\alpha0}\left(\overrightarrow{k}-i\overrightarrow{K}_{N}\right)\label{eq:4.3.1}\end{equation}

From the dispersion relation (\ref{eq:4.3.7},\ref{eq:4.3.8},\ref{eq:4.3.1})
it becomes clear that in the first approximation the presence of gradients
$\overrightarrow{K}_{N}$ of electron density logarithm change decrement
(imaginary part of $\frac{\partial\psi}{\partial t}$), changes effective
collision frequencies for ions with neutrals and makes them anisotropic
at high altitudes. All these changes are proportional to the scalar
product of the gradient of the electron density logarithm and average
electron velocity. Actually this fact contradicts with current theories
suggesting that in most cases only the electron density gradients
perpendicular to the magnetic field must be taken into account \cite{Kelley1989},
since there could be conditions when $\overrightarrow{V}_{\alpha0}$
is not perpendicular to the magnetic field, for example in case of
non-perpendicular magnetic and electric fields.

\subsection{Low-frequency solution}

\subsubsection{The solution nearest to zero}

The dispersion relation (\ref{eq:4.3.7}) has 6 solutions, and in
basic case all the solutions can be found only numerically. Lets find
the simplest approximate solution - nearest to zero. The solution
nearest to zero has a clear physical sence: in absence of average
plasma drifts and gradients plasma can be supposed as static and irregularities
should be static, i.e. 

\begin{equation}
\frac{\partial\psi}{\partial t}=x_{0}=0\label{eq:4_.3.15}\end{equation}

In presence of weak drifts and gradients we can suppose that the solution
is close to zero.

To find the solution nearest to zero we will use zero order Newton
solution (see, for example \cite{Yamamoto2000}):

\begin{equation}
x_{0}=-\left.\frac{f(x)}{\left(\frac{df}{dx}\right)}\right|_{x=0}\label{eq:4_.3.16}\end{equation}

By substituting the basic relations:

\begin{equation}
\begin{array}{l}
f(0)=m_{\beta}\left(\delta\widetilde{\omega}_{\beta\alpha N}\right)\left(\delta\widetilde{\omega}_{\beta\alpha N}-i\nu_{\beta nN}^{t,\mu}\right)\cdot\\
\cdot\left(\Omega_{\beta}^{2}+\left(\nu_{\beta nN}^{t,\mu}+i\left(\delta\widetilde{\omega}_{\beta\alpha N}\right)\right)^{2}\right)\cdot\\
\cdot\left(\Omega_{\alpha}^{2}\left(\widehat{b}\overrightarrow{k}\right)^{2}+\left(\nu_{\alpha nN}^{t,\mu}\right)^{2}\left(\overrightarrow{k}\right)^{2}\right)+\\
-\left(T_{\alpha}+T_{\beta}\right)\left(\left|\widehat{b}\overrightarrow{k}\right|^{2}\Omega_{\alpha}^{2}+\left|\overrightarrow{k}\right|^{2}\left(\nu_{\alpha nN}^{t,\mu}\right)^{2}\right)\cdot\\
\cdot\left(\Omega_{\beta}^{2}\left(\widehat{b}\overrightarrow{k}\right)^{2}+\left(\nu_{\beta nN}^{t,\mu}+i\left(\delta\widetilde{\omega}_{\beta\alpha N}\right)\right)^{2}\left(\overrightarrow{k}\right)^{2}\right)\end{array}\label{eq:4_.3.7}\end{equation}

\begin{equation}
\begin{array}{l}
\left(\frac{df}{dx}\right)_{x=0}=m_{\alpha}\left(-i\nu_{\alpha nN}^{t,\mu}\right)\left(\Omega_{\alpha}^{2}+\left(\nu_{\alpha nN}^{t,\mu}\right)^{2}\right)\cdot\\
\cdot\left(\Omega_{\beta}^{2}\left(\widehat{b}\overrightarrow{k}\right)^{2}+\left(\nu_{\beta nN}^{t,\mu}+i\delta\widetilde{\omega}_{\beta\alpha N}\right)^{2}\left(\overrightarrow{k}\right)^{2}\right)+\\
+m_{\beta}\left(2\delta\widetilde{\omega}_{\beta\alpha N}-i\nu_{\beta nN}^{t,\mu}\right)\cdot\left(\Omega_{\beta}^{2}+\left(\nu_{\beta nN}^{t,\mu}+i\delta\widetilde{\omega}_{\beta\alpha N}\right)^{2}\right)\cdot\\
\cdot\left(\Omega_{\alpha}^{2}\left(\widehat{b}\overrightarrow{k}\right)^{2}+\left(\nu_{\alpha nN}^{t,\mu}\right)^{2}\left(\overrightarrow{k}\right)^{2}\right)+\\
+2im_{\beta}\left(\delta\widetilde{\omega}_{\beta\alpha N}\right)\left(\delta\widetilde{\omega}_{\beta\alpha N}-i\nu_{\beta nN}^{t,\mu}\right)\cdot\\
\cdot\left(\left(\nu_{\beta nN}^{t,\mu}+i\delta\widetilde{\omega}_{\beta\alpha N}\right)\left(\Omega_{\alpha}^{2}\left(\widehat{b}\overrightarrow{k}\right)^{2}+\left(\nu_{\alpha nN}^{t,\mu}\right)^{2}\left(\overrightarrow{k}\right)^{2}\right)\right.+\\
\left.+\left(\Omega_{\beta}^{2}+\left(\nu_{\beta nN}^{t,\mu}+i\delta\widetilde{\omega}_{\beta\alpha N}\right)^{2}\right)\nu_{\alpha nN}^{t,\mu}\left(\overrightarrow{k}\right)^{2}\right)+\\
-2i\left(T_{\alpha}+T_{\beta}\right)\left|\overrightarrow{k}\right|^{2}\cdot\\
\cdot\left(\nu_{\alpha nN}^{t,\mu}\left(\Omega_{\beta}^{2}\left(\widehat{b}\overrightarrow{k}\right)^{2}+\left(\nu_{\beta nN}^{t,\mu}+i\delta\widetilde{\omega}_{\beta\alpha N}\right)^{2}\left(\overrightarrow{k}\right)^{2}\right)\right.\\
\left.+\left(\left(\left|\widehat{b}\overrightarrow{k}\right|^{2}\Omega_{\alpha}^{2}+\left|\overrightarrow{k}\right|^{2}\left(\nu_{\alpha nN}^{t,\mu}\right)^{2}\right)\left(\nu_{\beta nN}^{t,\mu}+i\delta\widetilde{\omega}_{\beta\alpha N}\right)\right)\right)\end{array}\label{eq:4_.3.8}\end{equation}

The low-frequency branch (\ref{eq:4_.3.16},\ref{eq:4_.3.7},\ref{eq:4_.3.8})
is pretty complex so we will investigate it at different ionospheric
heights.

\subsubsection{Obtaining traditional solution at 80-120km heights}

Let us analyze the branch (\ref{eq:4_.3.16},\ref{eq:4_.3.7},\ref{eq:4_.3.8})
for the typical ionospheric heights 80-120km. Within standard for
E-layer assumptions of $x,\delta\widetilde{\omega}_{ieN}<<\nu_{enN}^{t,\mu},\nu_{inN}^{t,\mu}$
, magnetized electrons and unmagnetized ions, and neglecting $\delta\widetilde{\omega}_{ieN}<<\nu_{inN}^{t,\mu}$,
we obtain following equations for function and its first differential
(neglecting in first differential by all the terms, proportional to
$\delta\widetilde{\omega}_{ieN}$ or $T_{e}+T_{i}$, based on suggestion
that Doppler shifts for ionacoustic or average velocities are sufficiently
small in comparison with $\nu_{inN}^{t,\mu}$):

\begin{equation}
\begin{array}{l}
f(0)=\left(m_{i}\delta\widetilde{\omega}_{ieN}\left(\delta\widetilde{\omega}_{ieN}-i\nu_{inN}^{t,\mu}\right)-\left(T_{e}+T_{i}\right)\left(\overrightarrow{k}\right)^{2}\right)\cdot\\
\cdot\left(\Omega_{e}^{2}\left(\widehat{b}\overrightarrow{k}\right)^{2}+\left(\nu_{enN}^{t,\mu}\right)^{2}\left(\overrightarrow{k}\right)^{2}\right)\left(\nu_{inN}^{t,\mu}\right)^{2}\end{array}\label{eq:5.1}\end{equation}

\begin{equation}
\begin{array}{l}
\left.\frac{df(x)}{dx}\right|_{x=0}=-im_{i}\nu_{enN}^{t,\mu}\Omega_{e}\Omega_{i}\left(\nu_{inN}^{t,\mu}\right)^{2}\left(\overrightarrow{k}\right)^{2}+\\
-im_{i}\left(\nu_{inN}^{t,\mu}\right)\left(\nu_{inN}^{t,\mu}\right)^{2}\left(\Omega_{e}^{2}\left(\widehat{b}\overrightarrow{k}\right)^{2}+\left(\nu_{enN}^{t,\mu}\right)^{2}\left(\overrightarrow{k}\right)^{2}\right)\end{array}\label{eq:5.3}\end{equation}

From (\ref{eq:5.3},\ref{eq:5.1}) and Newton method (\ref{eq:4_.3.16})
we obtain the solution, nearest to zero:

\begin{equation}
x_{0}=\frac{-\delta\widetilde{\omega}_{ieN}-i\frac{1}{\nu_{inN}^{t,\mu}}\left(\left(\delta\widetilde{\omega}_{ieN}\right)^{2}-\left(\frac{T_{e}+T_{i}}{m_{i}}\right)\left(\overrightarrow{k}\right)^{2}\right)}{1+\frac{\Omega_{e}\Omega_{i}}{\left(\nu_{inN}^{t,\mu}\right)\nu_{enN}^{t,\mu}}\frac{\left(\overrightarrow{k}\right)^{2}\left(\nu_{enN}^{t,\mu}\right)^{2}}{\left(\left|\widehat{b}\overrightarrow{k}\right|^{2}\Omega_{e}^{2}+\left|\overrightarrow{k}\right|^{2}\left(\nu_{enN}^{t,\mu}\right)^{2}\right)}}\label{eq:5.5}\end{equation}

Considering (\ref{eq:4.3.1}) we obtain the following solution for
plasma irregularities:

\begin{equation}
\begin{array}{l}
\frac{\partial\psi}{\partial t}=\frac{\left(\overrightarrow{V}_{e0}+\Psi\overrightarrow{V}_{i0}\right)\overrightarrow{k}}{\Psi+1}+\\
\,-i\frac{\frac{\Psi}{\nu_{inN}^{t,\mu}}\left(\left(\left(\overrightarrow{V}_{e0}-\overrightarrow{V}_{i0}\right)\overrightarrow{k}\right)^{2}-\left(\frac{T_{e}+T_{i}}{m_{i}}\right)\left(\overrightarrow{k}\right)^{2}\right)}{\Psi+1}\\
-i\frac{\left(\overrightarrow{V}_{e0}+\Psi\overrightarrow{V}_{i0}\right)\overrightarrow{K}_{N}}{\Psi+1}\end{array}\label{eq:5.6}\end{equation}

\begin{equation}
\Psi=\frac{\nu_{inN}^{t,\mu}\nu_{enN}^{t,\mu}}{\Omega_{e}\Omega_{i}}\frac{\left|\widehat{b}\overrightarrow{k}\right|^{2}\Omega_{e}^{2}+\left|\overrightarrow{k}\right|^{2}\left(\nu_{enN}^{t,\mu}\right)^{2}}{\left(\overrightarrow{k}\right)^{2}\left(\nu_{enN}^{t,\mu}\right)^{2}}\label{eq:5.7}\end{equation}

In gradient-free case the solution (\ref{eq:5.6}) looks exactly as
the standard one \cite{Kelley1989}, in presence of gradients the
solution differs from the standard one, most probably due to custom
direction of gradients and custom orientation of velocities.

\subsubsection{Fully magnetized case or instabilities at altitudes above 140km }

Let us analyze the branch (\ref{eq:4_.3.16},\ref{eq:4_.3.7},\ref{eq:4_.3.8})
in case of sufficiently high altitudes, when both types of charged
particles are magnetized (i.e. from about 130-140 km). At high altitudes
we can neglect the difference in electron and ion velocities in comparison
with their absolute values (both components are magnetized and move
with almost the same velocities):

\begin{equation}
|Re(x)|>>|\delta\widetilde{\omega}_{\beta\alpha N}|\label{eq:4.3.10}\end{equation}

Supposing $\delta\widetilde{\omega}_{\beta\alpha N}=0$, and when 
investigating wavevectors perpendicular to the magnetic field,
the solution becomes simplier:

\begin{equation}
\begin{array}{l}
f(0)=-\left(T_{\alpha}+T_{\beta}\right)\left|\overrightarrow{k}\right|^{4}\left(\nu_{\alpha nN}^{t,\mu}\right)^{2}\left(\nu_{\beta nN}^{t,\mu}\right)^{2}\end{array}\label{eq:4__1.3.7}\end{equation}

\begin{equation}
\begin{array}{l}
\left(\frac{df}{dx}\right)_{x=0}=m_{\alpha}\left(-i\nu_{\alpha nN}^{t,\mu}\right)\left(\Omega_{\alpha}^{2}+\left(\nu_{\alpha nN}^{t,\mu}\right)^{2}\right)\left(\nu_{\beta nN}^{t,\mu}\right)^{2}\left(\overrightarrow{k}\right)^{2}+\\
+m_{\beta}\left(-i\nu_{\beta nN}^{t,\mu}\right)\left(\Omega_{\beta}^{2}+\left(\nu_{\beta nN}^{t,\mu}\right)^{2}\right)\left(\nu_{\alpha nN}^{t,\mu}\right)^{2}\left(\overrightarrow{k}\right)^{2}+\\
-2i\left(T_{\alpha}+T_{\beta}\right)\left|\overrightarrow{k}\right|^{4}\nu_{\beta nN}^{t,\mu}\nu_{\alpha nN}^{t,\mu}\left(\nu_{\beta nN}^{t,\mu}+\nu_{\alpha nN}^{t,\mu}\right)\end{array}\label{eq:4__1.3.8}\end{equation}

After simple arithmetic and by taking into account magnetized plasma
and typical ionospheric conditions ($\alpha$- electrons):

\[
\Omega_{\alpha}^{2}>>\left(\nu_{\alpha nN}^{t,\mu}\right)^{2};\Omega_{\beta}^{2}>>\left(\nu_{\beta nN}^{t,\mu}\right)^{2};\nu_{\beta nN}^{t,\mu}<<\nu_{\alpha nN}^{t,\mu}\]

we have:

\begin{equation}
\begin{array}{l}
f(0)=-\left(T_{\alpha}+T_{\beta}\right)\left|\overrightarrow{k}\right|^{4}\left(\nu_{\alpha nN}^{t,\mu}\right)^{2}\left(\nu_{\beta nN}^{t,\mu}\right)^{2}\end{array}\label{eq:4_.3.7b}\end{equation}

\begin{equation}
\begin{array}{l}
\left(\frac{df}{dx}\right)_{x=0}=m_{\alpha}\left(-i\nu_{\alpha nN}^{t,\mu}\right)\left(\Omega_{\alpha}^{2}\right)\left(\nu_{\beta nN}^{t,\mu}\right)^{2}\left(\overrightarrow{k}\right)^{2}+\\
+m_{\beta}\left(-i\nu_{\beta nN}^{t,\mu}\right)\left(\Omega_{\beta}^{2}\right)\left(\nu_{\alpha nN}^{t,\mu}\right)^{2}\left(\overrightarrow{k}\right)^{2}+\\
-2i\left(T_{\alpha}+T_{\beta}\right)\left|\overrightarrow{k}\right|^{4}\nu_{\beta nN}^{t,\mu}\left(\nu_{\alpha nN}^{t,\mu}\right)^{2}\end{array}\label{eq:4_.3.8b}\end{equation}

Taking into account the typical ionospheric conditions:

\begin{equation}
m_{e}\Omega_{e}^{2}\nu_{inN}^{t,\mu}>>m_{i}\Omega_{i}^{2}\nu_{enN}^{t,\mu},\left(T_{i}+T_{e}\right)\left|\overrightarrow{k}\right|^{2}\nu_{enN}^{t,\mu}\label{eq:4.3.19a}\end{equation}

\begin{equation}
x_{0}\approx i\frac{\nu_{enN}^{t,\mu}\left(T_{e}+T_{i}\right)\left|\overrightarrow{k}\right|^{2}}{\left(\Omega_{e}^{2}\right)m_{e}}\label{eq:4.3.19}\end{equation}

\begin{equation}
\frac{\partial\psi}{\partial t}\approx i\frac{\nu_{enN}^{t,\mu}\left(T_{e}+T_{i}\right)\left|\overrightarrow{k}\right|^{2}}{\left(\Omega_{e}^{2}\right)m_{e}}-i\overrightarrow{K}_{N}\overrightarrow{V}_{e0}+\overrightarrow{V}_{e0}\overrightarrow{k}\label{eq:4.3.20}\end{equation}

From (\ref{eq:4.3.20}) the condition for the growing solution $Im\left(\frac{\partial\psi}{\partial t}\right)<0$
becomes:

\begin{equation}
\overrightarrow{K}_{N}\overrightarrow{V}_{e0}>D_{A\bot}\left|\overrightarrow{k}\right|^{2}\label{eq:4.3.21}\end{equation}

where

\begin{equation}
D_{A\bot}=\frac{\nu_{enN}^{t,\mu}\frac{\left(T_{e}+T_{i}\right)}{m_{i}}}{\Omega_{e}\Omega_{i}}\label{eq:4.3.21a}\end{equation}
 is the so called coefficient of ambipolar diffusion \cite{Galant1977}.

For typical ionospheric conditions the growth condition (\ref{eq:4.3.21})
can be estimated as:

\begin{equation}
\overrightarrow{K}_{N}\overrightarrow{V}_{e0}>0.2[m^{2}/s]\left|\overrightarrow{k}\right|^{2}\label{eq:4.3.22}\end{equation}

The spectral offset for these irregularities (\ref{eq:4.3.20}) is
exactly the Doppler drift in crossed fields (and defined by zero-order
solution (\ref{eq:3.2.1}-\ref{eq:3.2.2})):

\begin{equation}
Re\left(\frac{\partial\psi}{\partial t}\right)\approx\overrightarrow{V}_{e0}\overrightarrow{k}\label{eq:4.3.23}\end{equation}

It is necessary to note that the solution is obtained in weak gradients
approximation (\ref{eq:4.2.18e}) valid when:

\begin{equation}
\left|\overrightarrow{k}\right|\frac{\nu_{enN}^{t,\mu}}{\Omega_{e}}>>\left|\overrightarrow{K}_{N}\right|\label{eq:4.3.23b}\end{equation}

It should be noted that possible relation of the ambipolar diffusion
with irregularities existence at these heights has been noted in \cite{KaganKelley2000},
but the problem was not investigated in detail. The close condition
for irregularities growth at high altitudes $V_{e}>\eta D_{A\bot}$
was also obtained by \cite{Gershman1974}, but with another proportionality
coefficient $\eta$ and in qualitative analysis of a model case.

Due to the solution (\ref{eq:4.3.20}) corresponds to the same branch
(\ref{eq:4_.3.16},\ref{eq:4_.3.7},\ref{eq:4_.3.8})
as well-known gradint-drift instabilities (\ref{eq:5.6},\ref{eq:5.7})
, below we will call this kind of solition as 'fully magnetized gradient-drift
instabilities' (FMGD) to stress that this is the same gradient-drift
branch but in a bit different conditions. 

Starting from early 1960s \cite{Bowles1962,Blasley1964} at equatorial
HF and UHF radars researchers observe an unique type of echo, the
so called 150km equatorial one. There are some theories to explain
it (for example \cite{KudekiFawcett1993,Tsunoda1994,TsunodaEcklund2004,KaganKelley2000,CosgroveTsunoda2002})
, but the exact physical mechanism of it is still unclear \cite{ChoudharyStMauriceMahajan2004,ChauKudeki2006,ChauEtAl2009}.
The geometry at equator (horizontal magnetic field, almost upward
drift velocity) allows us to use standard upward vertical gradient
as a source for generation of this kind of instabilities. For sounding
frequency of equatorial radar Jicamarca ($k\sim0.3$) and for $K_{N}\sim10^{-3}\div10^{-4}[m^{-1}]$
(standard vertical electron density gradients) the growth condition
(\ref{eq:4.3.21}) becomes:

\begin{equation}
\overrightarrow{V}_{e0}>20\div200[m/s]\label{eq:4.3.23.1}\end{equation}

It should be noted that for gradients higher than (\ref{eq:4.3.23b})
($K_{N}>10^{-3}$) one should take into account all the terms in (\ref{eq:4.2.17})
instead of using only (\ref{eq:4.2.18}), but this should not affect
too much the observed effect.

To analyze the properties of the echo, some modelling has been done
using the latest Internation Refference Ionosphere (IRI-2007) model.
The height and time dependence of the electron density gradients are
most important for the generation of this type of irregularities.
We have analyzed 13 years period (1990-2002) using IRI model (for
typical non-disturbed conditions $f_{10.7}=150,\, A_{p}=10$) and
obtained the following results.

At Fig.1 the altitudinal dependence of $K_{N}$ is shown. Points are
the hourly values over the whole period of 13 years. As one can see,
there is a maximum $K_{N}>10^{-4}$ at heights 135-180km. So, this
kind of irregularities could arise at heights 135-180km, and it corresponds
well with the experimental observations statistics \cite{ChauKudeki2006,ChoudharyStMauriceMahajan2004}.

At Fig.2 an hourly dependence of the $K_{N}$ is shown, as a function
of UT for heights 140-200km. As one can see, the time dependence of
the gradients has a most intensive maximum between 14:00 and 19:00
UT (9:00-14:00LT). This also corresponds well with the experimental
observations \cite{ChauKudeki2006}.

The dependence of irregularities frequency (\ref{eq:4.3.23}) corresponds
well with the empirical models \cite{ChauWoodman2004,RaghavaraEtAl2002}
and allows to interpret the experimental data as Doppler frequency
offset due to electron drift in crossed fields.

According to the experimental observations, the echo starts with $V_{e}>10m/s$,
according to our calculations it should start with $V_{e}>20\div200m/s$.
One of the mechanisms allowing to lower the speed limit was suggested
in \cite{KudekiFawcett1993}. They suggest that acoustic-gravitational
waves can be responsible for the triggering the instabilities. In
our terms, the acoustic-gravitational waves will produce gradients
more than $K_{N}>10^{-3}$, and this will produce this type of irregularities
even at lower velocities, for example at $V_{e}>10m/s$. Another possible
mechanism that will lower the velocities necessary for generation
of the instability is an observation of high step-like gradients at
these heights from the rocket data (see for example \cite{RaghavaraEtAl2002}).
They should also produce the increase of $K_{N}$ high enough for
lowering the speed limit.

Summarizing all said above we can suggest that the FMGD instabilities
can be the source of 150km equatorial echo and this theory can be
used for experiment interpretation.

\section{Conclusions}

In the paper within the approximation of the two-fluid magnetohydrodynamics
and geometrooptical approximation the dispersion relation (\ref{eq:4.3.7},
\ref{eq:4.3.8}, \ref{eq:4.2.2}-\ref{eq:4.2.5}) at 80-200km altitudes
was obtained. The relation describes ionacoustic instabilities of
the ionospheric plasma at 80-200km altitudes in three-dimensional
weakly irregular ionosphere.

It was shown that not only electron density gradients perpendicular
to the magnetic field should be taken into account when investiagting
ionospheric instabilities, but gradients along the average drift velocity
(\ref{eq:4.3.7}, \ref{eq:4.3.8}, \ref{eq:4.2.3}, \ref{eq:4.2.5}).

The dispersion relation obtained has a form of the 6-th order polynomial
for the oscillation frequency. 

It is shown, that a solution branch exists that grows with time and
describe instabilities both at 80-120km heights and 135-180km heights.

For altitudes 80-120km the solution close to the standard one (\ref{eq:5.6},
\ref{eq:5.7}) and corresponds to the Farley-Buneman and gradient-drift
instabilties.

The difference between obtained (\ref{eq:4_.3.16},\ref{eq:4_.3.7},\ref{eq:4_.3.8})
and standard solutions \cite{Kelley1989} becomes significant at altitudes
above 140 km, where standard one is not valid. As the analysis shown
at these altitudes the solution grows with time (\ref{eq:4.3.20},
\ref{eq:4.3.21}). The conditions for the growth is the presence of
co-directed electron density gradients and electron drifts and perpendicularity
of line-of-sight to the magnetic field. These conditions are regularly
satisfied at magnetic equator for expected conditions (\ref{eq:4.3.23.1}).
Detailed analysis has shown that this solution could explain a lot
of properties of 150 km equatorial radioecho - the ionospheric phenomena
that has no explanation for more than 40 years.
%% \section{Acknowledgements}
\ack
Authors thanks to N.Nishitani for fruitful discussion. The work was
done under financial support of RFBR grant \#07-05-01084a.

\begin{table}[t]

\caption{Thermal velosities, hyrofrequencies and frequencies of collisions
with neutrals for typical ionospheric conditions. Frequencies in Hz,
Velocities in m/sec, height in km}

\begin{tabular}{|c|c|c|c|c|c|c|}
\hline 
$h$ &
$\sqrt{\frac{T_{e}}{m_{e}}}$ &
$\Omega_{e}$&
$\nu_{en}^{t,\mu}$ &
$\sqrt{\frac{T_{i}}{m_{i}}}$ &
$\Omega_{i}$ &
$\nu_{in}^{t,\mu}$ \tabularnewline
\hline
\hline 
80&
6e+4&
1e+7&
3e+7&
3e+2&
2e+2&
2e+5\tabularnewline
\hline 
100&
5e+4&
1e+7&
6e+5&
2e+2&
2e+2&
4e+3\tabularnewline
\hline 
120&
8e+4&
1e+7&
4e+4&
3e+2&
2e+2&
2e+2\tabularnewline
\hline 
140&
1e+5&
1e+7&
1e+4&
4e+2&
2e+2&
4e+1\tabularnewline
\hline 
160&
1e+5&
9e+6&
5e+3&
5e+2&
2e+2&
2e+1\tabularnewline
\hline 
180&
1e+5&
9e+6&
3e+3&
6e+2&
2e+2&
8\tabularnewline
\hline 
200&
1e+5&
9e+6&
2e+3&
7e+2&
2e+2&
4\tabularnewline
\hline
\end{tabular}
\end{table}

\begin{figure}[t]
\vspace*{2mm}

\includegraphics{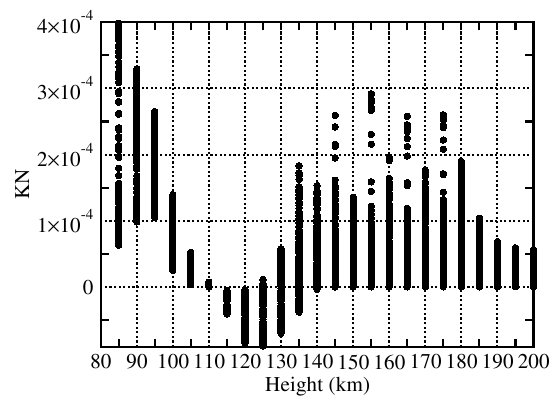}

\caption{$K_{N}$ dependence on height over the 1990-2002 years, points are
the hourly values.}
\end{figure}

\begin{figure}[t]
\vspace*{2mm}

\includegraphics{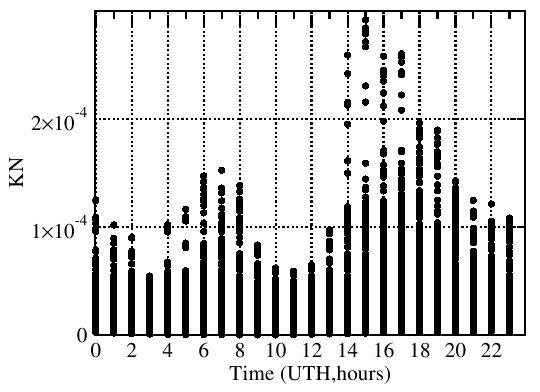}

\caption{$K_{N}$ dependence on time at 140-200km over the 1990-2002 years,
points are the hourly values.}
\end{figure}

\appendix

\section{Approximations used}

The theory limitations are listed, mostly following to the \cite{Gurevich1973,Galant1977,Dimant1995}.

At the altitudes 80-200km we suggest that the following conditions
are satisfied:

$\tau^{-1}<<\Omega_{i}$ - all the basic plasma parameters has only
slow variations and plasma supposed to be quasistatic;

$\delta_{en}<<1$ - average loss of energy of electrons with neutrals
is small enough;

$\nu_{in}<<\nu_{en}$;$\nu_{in}<<\Omega_{e}$ ;

$\nu_{ii}<<\nu_{in}$ - ion-ion collisions are rare enough to take
into account only ion-neutral collisions. Not valid above 200km.

$\nu_{ei}\sim\nu_{ee}<<\nu_{en}$ - electron-ion and electron-electron
collisions are rare enough to take into account only electron-neutral
collisions. Not valid above 200km.

$k_{\bot}\rho_{e}<<1$ electron hyroradius much smaller than wavelength.

$\rho_{e}<<\lambda_{e}<<k_{||}^{-1}<<L_{||}$ - plasma is quasihomogeneous
enough for GO approximation to be valid.

$\lambda_{d}k<<1$ - wavelength is much bigger than Debye radius.

$\delta_{ei}\nu_{ei}<<\nu_{in}$ - necessary for independent thermalization
of ions and electrons, in this approximation the average collision
frequency does not depend on particles velocity or motion direction
\cite{Gurevich1973}.

$V_{0n}=0$ - average speed of neutrals is much smaller than electrons
and ions speed. Allows us to neglect neutral motions.

In more details the MHD validity conditions can be found in \cite{Gurevich1973,Galant1977}.

\section{Inversion of the matrix operator $\widehat{P}_{\alpha,0}$ (IAQV
approximation)}

Lets analyze inversion of the matrix operator $\widehat{P}_{\alpha,0}$
(\ref{eq:3.3.7}-\ref{eq:3.3.10}). One can see, that in special case
$\widehat{P}_{7}=0$ the inversion is very easy. In this case by taking
into account that $\overrightarrow{P}_{6}$ is static, we can create
the coordinate system, based on unity vector $\widehat{b}=\overrightarrow{P}_{6}/|\overrightarrow{P}_{6}|$,
which is antiparallel to the magnetic field. In this case we can write:

\begin{equation}
\widehat{P}_{0}\overrightarrow{f}=P_{5}\overrightarrow{f}_{||}+P_{5}\overrightarrow{f}_{\bot}+P_{6}\overrightarrow{f}_{\bot}\times\widehat{b}\label{eq:p1}\end{equation}

Where $||,\bot$ means parallel and perpendicular to the $\widehat{b}$.

By making scalar and vector products of (\ref{eq:p1}) with $\widehat{b}$
we have:

\begin{equation}
\left\{ \begin{array}{l}
\widehat{b}(\widehat{P}_{0}\overrightarrow{f})=P_{5}\widehat{b}\overrightarrow{f}_{||}\\
\widehat{b}\times(\widehat{P}_{0}\overrightarrow{f})=P_{5}(\widehat{b}\times\overrightarrow{f}_{\bot})+P_{6}\widehat{b}\times(\overrightarrow{f}_{\bot}\times\widehat{b})\end{array}\right.\label{eq:p2}\end{equation}

From first equation (\ref{eq:p2}):

\begin{equation}
\overrightarrow{f}_{||}=\widehat{b}\frac{\widehat{b}(\widehat{P}_{0}\overrightarrow{f})}{P_{5}}\label{eq:p3}\end{equation}

After comparing the second equation in (\ref{eq:p2}) and its vector
product with $\widehat{b}$ we have:

\begin{equation}
\left\{ \begin{array}{l}
-\frac{\widehat{b}\times(\widehat{P}_{0}\overrightarrow{f})-P_{5}(\widehat{b}\times\overrightarrow{f}_{\bot})}{P_{6}}=\widehat{b}\times(\widehat{b}\times\overrightarrow{f}_{\bot})\\
\frac{\widehat{b}\times(\widehat{b}\times(\widehat{P}_{0}\overrightarrow{f}))-P_{6}\widehat{b}\times(\widehat{b}\times(\overrightarrow{f}_{\bot}\times\widehat{b}))}{P_{5}}=\widehat{b}\times(\widehat{b}\times\overrightarrow{f}_{\bot})\end{array}\right.\label{eq:p4}\end{equation}

Therefore, by taking into account the properties of double vector
product:

\begin{equation}
\begin{array}{l}
-\frac{\widehat{b}\times(\widehat{P}_{0}\overrightarrow{f})-P_{5}(\widehat{b}\times\overrightarrow{f}_{\bot})}{P_{6}}=\\
\,\,\,=\frac{\widehat{b}\times(\widehat{b}\times(\widehat{P}_{0}\overrightarrow{f}))-P_{6}(\widehat{b}\times\overrightarrow{f}_{\bot})}{P_{5}}\end{array}\label{eq:p7}\end{equation}

So

\begin{equation}
(\widehat{b}\times\overrightarrow{f}_{\bot})=\frac{P_{5}\widehat{b}\times(\widehat{P}_{0}\overrightarrow{f})+P_{6}(\widehat{b}\times(\widehat{b}\times(\widehat{P}_{0}\overrightarrow{f})))}{P_{5}^{2}+P_{6}^{2}}\label{eq:p8}\end{equation}

and (after making vector product with $\widehat{b}$ and some vector
algebra):

\begin{equation}
\overrightarrow{f}_{\bot}=\frac{P_{6}(\widehat{b}\times(\widehat{P}_{0}\overrightarrow{f}))-P_{5}\widehat{b}\times(\widehat{b}\times(\widehat{P}_{0}\overrightarrow{f}))}{P_{6}^{2}+P_{5}^{2}}\label{eq:p9}\end{equation}

Summarizing (\ref{eq:p3}) and (\ref{eq:p9}) we have $\overrightarrow{f}=\overrightarrow{f}_{\bot}+\overrightarrow{f}_{||}$:

\begin{eqnarray}
\overrightarrow{f} & = & \frac{P_{6}P_{5}(\widehat{b}\times(\widehat{P}_{0}\overrightarrow{f}))-P_{5}^{2}\widehat{b}\times(\widehat{b}\times(\widehat{P}_{0}\overrightarrow{f}))}{\left(P_{6}^{2}+P_{5}^{2}\right)P_{5}}+\label{eq:p10}\\
 & + & \frac{\left(P_{6}^{2}+P_{5}^{2}\right)\widehat{b}\left(\widehat{b}(\widehat{P}_{0}\overrightarrow{f})\right)}{\left(P_{6}^{2}+P_{5}^{2}\right)P_{5}}\nonumber \end{eqnarray}

or

\begin{equation}
\widehat{P}_{0}^{-1}\overrightarrow{f}=\frac{\widehat{b}\left(\widehat{b}\overrightarrow{f}\right)P_{6}^{2}+P_{5}^{2}\overrightarrow{f}+P_{5}P_{6}(\widehat{b}\times\overrightarrow{f})}{P_{5}\left(R_{6}^{2}+R_{5}^{2}\right)}\label{eq:p12}\end{equation}

It is clear that this approximation is valid when:

\begin{equation}
\left|P_{\alpha5}\overrightarrow{f}+\overrightarrow{f}\times\overrightarrow{P_{\alpha6}}\right|>>\left|\widehat{P_{\alpha7}}\overrightarrow{f}\right|\label{eq:p13}\end{equation}

Qualitatively one can estimate the orders of terms:

\begin{equation}
\nu_{\alpha n}^{t,\mu},\Omega_{\alpha}>>V_{\alpha0}L_{V}^{-1}\label{eq:p14}\end{equation}

where

\begin{equation}
L_{V}^{-1}=\frac{\left|\frac{\partial V_{\alpha0(j)}}{\partial r_{(k)}}\right|}{V_{\alpha0}}\label{eq:p15}\end{equation}

For maximal ionospheric disturbances up to 200km height we can estimate
$V_{\alpha0}<3000[m/s]$, $\nu_{en}^{t,\mu}>100Hz$, $\nu_{in}^{t,\mu}>10Hz$,
$\Omega_{\alpha}>100Hz$. In this case the validity condition has
the form

\begin{equation}
L_{V}>>300[m]\label{eq:p20}\end{equation}

Summarizing, in very disturbed ionosphere the characteristic changes
of parameters should not exceed couple hundreds meters, for less disturbed
conditions these limitations becomes even weaker. So the obtained
approximation for $\widehat{P}_{\alpha,0}^{-1}$ (\ref{eq:p12}) is
valid for most part of cases below 200km.

\newpage{}

\end{document}